\documentclass[twocolumn,preprintnumbers,prl,superscriptaddress]{revtex4-1}

\usepackage{amsthm,amsmath,amssymb,physics,mathtools}
\usepackage{bbm}

\usepackage{hyperref}
\hypersetup{
    colorlinks = true,       
    linkcolor = red,          
    citecolor = magenta,        
    filecolor=red,      
    urlcolor= cyan           
}

\usepackage{graphicx}
\usepackage{xcolor}

\newcommand{\be}{\begin{equation}}
\newcommand{\ee}{\end{equation}}
\newcommand{\beq}{\begin{equation}}
\newcommand{\eeq}{\end{equation}}
\newcommand{\bea}{\begin{equation} \begin{aligned}}
\newcommand{\eea}{ \end{aligned} \end{equation}}
\newcommand{\per}{\, .}

\newcommand{\ii}{ \mathrm{i}}
\newcommand{\Y}{\mathrm{Y}}
\newcommand{\LL}{\mathrm{L}}

\newcommand{\K}{\mathrm{K}}
\newcommand{\delt}{\tilde{\delta}}
\newcommand{\tdelt}{\delta}

\begin{document}

\title{Thermodynamic Bethe Ansatz for Fishnet CFT}
\author{Benjamin Basso}
\email{benjamin.basso@ens.fr}
\affiliation{Laboratoire de physique de l'\'Ecole normale sup\'erieure, ENS, Universit\'e PSL, CNRS, Sorbonne Universit\'e, Universit\'e Paris-Diderot, Sorbonne Paris Cit\'e, 24 rue Lhomond, 75005 Paris, France}
\author{Gwena\"el Ferrando}
\email{gwenael.ferrando@ens.fr}
\affiliation{Laboratoire de physique de l'\'Ecole normale sup\'erieure, ENS, Universit\'e PSL, CNRS, Sorbonne Universit\'e, Universit\'e Paris-Diderot, Sorbonne Paris Cit\'e, 24 rue Lhomond, 75005 Paris, France}
\affiliation{Institut de Physique Th\'eorique, Universit\'e Paris-Saclay, CNRS, CEA, 91191 Gif-sur-Yvette, France}
\author{Vladimir Kazakov}
\email{vladimir.kazakov@ens.fr}
\affiliation{Laboratoire de physique de l'\'Ecole normale sup\'erieure, ENS, Universit\'e PSL, CNRS, Sorbonne Universit\'e, Universit\'e Paris-Diderot, Sorbonne Paris Cit\'e, 24 rue Lhomond, 75005 Paris, France}
\affiliation{Theoretical Physics Department, CERN, 1211 Geneva 23, Switzerland} 
\author{De-liang Zhong}
\email{zdlzdlzdl@gmail.com}
\affiliation{School of Physics and Astronomy, Tel Aviv University, Ramat Aviv 69978, Israel}

\date{\today}

\begin{abstract}
We present the TBA equations and the Y-system for the exact spectrum of general multi-magnon local operators in the $D$-dimensional anisotropic version of the bi-scalar fishnet CFT. The mixing matrix of such operators is given in terms of fishnet planar graphs of multi-wheel and multi-spiral type. These graphs probe the two main building blocks of the TBA approach that are the magnon dispersion relation and the magnon scattering matrix and which we both obtain by diagonalising suitable graph-building operators. We also obtain the dual version of the TBA equations, which relates, in the continuum limit,  $D$-dimensional graphs to two dimensional sigma models in $AdS_{D+1}$. It allows us to verify a general formula obtained by A.~Zamolodchikov for the critical coupling.

\end{abstract}

\maketitle

\section{\label{sec:intro}Introduction}

The fishnet conformal field theory (FCFT)~\cite{Gurdogan:2015csr}~(for a review see~\cite{Kazakov:2018hrh}) arises as a double scaling limit of weakly coupled and strongly \(\gamma\)-twisted \(\mathcal{N}=4\) SYM theory. It stands out as a striking example of a non-supersymmetric and yet integrable planar CFT in four dimensions, with an exactly marginal coupling \cite{Gromov:2018hut,Sieg:2016vap} and a non-trivial moduli space of vacua~\cite{Karananas:2019fox}. Due to these features, it has attracted a growing interest over the last few years~\cite{Caetano:2016ydc,Mamroud:2017uyz,Chicherin:2017cns,Chicherin:2017frs,Basso:2017jwq,Grabner:2017pgm,Gromov:2019aku,Gromov:2019bsj,Gromov:2019jfh,Gromov:2017cja,Korchemsky:2018hnb,Kazakov:2018gcy,Kazakov:2018hrh,Ipsen:2018fmu,deMelloKoch:2019ywq,Pittelli:2019ceq,Chowdhury:2019hns,Adamo:2019lor}. Moreover, unlike its supersymmetric parent, the theory can be defined in any dimension \(D\)~\cite{Kazakov:2018qbr,Zamolodchikov:1980mb}, called here FCFT\(_D\), with the Lagrangian
\bea\label{eqn-Lag}
\mathcal{L} = N_c \Tr& \Big[X^{\dagger}\left(-\partial_\mu\partial^\mu\right)^{\delt}X + Z^{\dagger}\left(-\partial_\mu\partial^\mu\right)^{\tdelt}Z \\ 
&+ (4\pi)^\frac{D}{2} \xi^2 X^{\dagger} Z^{\dagger} X Z \Big]\eea
where $X$ and $Z$ are two $N_c\times N_c$ matrix complex scalar fields with respective bare dimensions $\tdelt$ and $\delt = D/2 - \tdelt$ for $0<\tdelt<\frac{D}{2}$.

In the planar (large \(N_c\)) limit its perturbation expansion is dominated by conformal ``fishnet" Feynman graphs: at high order the bulk structure of such graphs has the shape of regular square lattice with $X$ and $Z$ propagators pointing in two orthogonal directions~\footnote{The structure of such graph near its boundary depends on the computed physical quantity}. First evidence for integrability came from A.~Zamolodchikov who treated the fishnet graphs with an appropriate choice of \(D\)-dimensional propagators as an integrable statistical mechanical  system~\cite{Zamolodchikov:1980mb}. The integrability of the fishnet graphs is also closely related to the integrability of conformal, non-compact  \(SO(1,D+1)\) spin chain with spins in principal series representations~\cite{Chicherin:2012yn}.
 
Quantum integrability of a QFT (defined, in a broad sense, as the existence of infinitely many conserved charges)  usually allows for a deep insight into its non-perturbative structure. It also provides us with the tools for some explicit (though not necessarily easy) calculations of basic physical quantities, such as the correlators of local operators. A remarkable progress in this direction has been achieved in the last 15 years in the most emblematic planar integrable CFT -- $\mathcal{N} = 4$ SYM \cite{Beisert:2010jr,Beisert:2005fw,Beisert:2006ez,Gromov:2009tv,Basso:2015zoa,Fleury:2017eph,Eden:2016xvg}. In particular, the computation of the spectrum of anomalous dimensions of local operators (encoded in the two-point functions) appeared to be possible via the thermodynamic Bethe ansatz (TBA)~\cite{Bombardelli:2009ns,Gromov:2009bc,Arutyunov:2009ur}, which finally evolved into the most efficient method of quantum spectral curve (QSC)~\cite{Gromov:2013pga,Gromov:2014caa}.

\begin{figure}[!ht]
    \begin{minipage}{0.24\textwidth}
    \centering
    \includegraphics[width=0.9\textwidth]{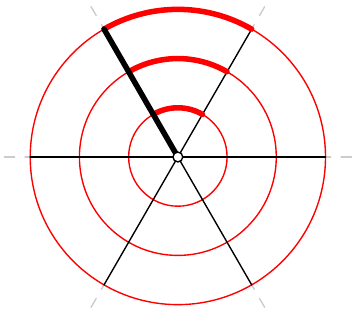}
    \end{minipage}\hfill
    \begin{minipage}{0.24\textwidth}
        \centering
        \includegraphics[width=0.9\textwidth]{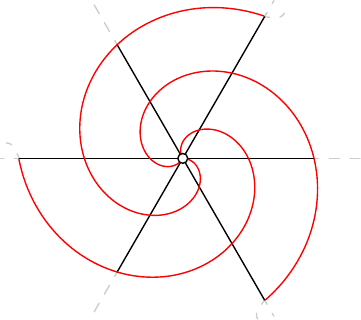} 
    \end{minipage}
    \caption{Specimens of planar fishnet graphs contributing to the anomalous dimensions of multi-magnon operators (central points) with black and red lines representing propagators of $Z$ and $X$ fields, respectively. Left panel: Multi-wheel graph renormalizing the ground-state operator with $M = 0$. The graph can be obtained by iterating the graph-building operator $\widehat{\Gamma}_{N=3}$ shown here in bold face. Right panel: Multi-spiral graph contributing to the mixing of excited-state operators with here $M =3$ magnons inserted at the origin.}
    \label{fig:graph}
\end{figure}

In this paper, we propose TBA equations for FCFT\(_{D}\), at any \(D\), for the dimensions of multi-magnon operators of the type
\beq\label{eq:oper}
   \mathcal{O}_{J,M}(x)=\Tr(X^M Z^J)+\ldots\,.
\eeq
The mixing matrix of such operators is entirely defined by multi-wheel or multi-spiral planar Feynman graphs~\cite{Gurdogan:2015csr,Caetano:2016ydc}, such as those on figure~\ref{fig:graph}. These integrals have attracted a considerable interest in the literature as examples of explicitly calculable  multi-loop Feynman graphs~\cite{Gromov:2017cja,Panzer:2015ida,Broadhurst:1985vq,Basso:2017jwq,Derkachov:2018rot}.

In the case of FCFT\(_{4}\), these TBA equations can be obtained by taking the double-scaling limit of the full TBA system of twisted $\mathcal{N}=4$ SYM \cite{Caetano:2016ydc,Gromov:2017cja, Ahn:2011xq}. We no longer have this luxury once we deal with FCFT\(_D\) with \(D\ne 4\), which does not have its SYM ``parent"~\footnote{For \(D=3\), the ABJM model, in a similar double scaling limit, becomes a FCFT dominated by regular triangular planar graphs~\cite{Caetano:2016ydc}}. In this case, to arrive at the TBA equations we shall rely on the direct fishnet graph computations as well as a certain intuition borrowed from \(D=4\) case. We will then derive the asymptotic Bethe ansatz~(ABA) equations, valid in the limit $J\to\infty$, and check them against the explicit fishnet-type Feynman integrals computations.

\section{Graph-building operators and scattering data}

The TBA construction relies on the knowledge of the asymptotic data, dispersion relation and factorised S-matrix, that characterise the integrable structure of the fishnet graphs. In planar $\mathcal{N}=4$ SYM these were determined using supersymmetry and crossing symmetry \cite{Beisert:2005tm,Janik:2006dc,Beisert:2006ib,Beisert:2006ez}. We cannot follow these steps for FCFT$_D$ for lack of symmetries but we can read off the scattering data from the graphs directly. In fact, the information can all be obtained from the wheel graphs shown on figure \ref{fig:graph} (left) and corresponding to the local operator \eqref{eq:oper} with $M=0$, referred to as vacuum state. General results for the excited states with $M\neq 0$ will be given in a subsequent section.

The S-matrix that is required here is the one controlling the scattering of magnons in the ``open string channel" aka mirror kinematics. The idea is to treat the \(X\) propagators along the angular direction in figure \ref{fig:graph} as magnon excitations moving radially along the \(Z\) propagators. Geometrically, this mirror 1-dimensional system emerges from the decomposition $\mathbb{R}^D\cong \mathbb{R}_+\times S^{D-1}$, with $r=e^{\sigma}\in \mathbb{R}_+$ being the distance to the origin, $\sigma$ the mirror position, and with the sphere $S^{D-1}$ giving rise to an internal $O(D)$ symmetry.

Mirror magnons evolve in this picture through the action of the graph-building operator
\beq
[\widehat{\Gamma}_N \Phi]({\bf x}) = \int \Phi({\bf y}) \prod_{i=1}^{N} \frac{\pi^{-D/2} \dd^D y_{i}}{(x_{i-1}-x_{ i})^{2\delt}(x_i-y_i)^{2\tdelt}}\, ,
\label{graph-building}
\eeq
which acts on $N$-magnon wave function $\Phi({\bf x}) = \Phi(x_1, \ldots,  x_N)$, with $x_0 = 0$.
Clearly, any wheel graph can be obtained by iteration of a graph-building operator, see figure \ref{fig:graph}. The significance of these operators in the fishnet theory was unveiled in \cite{Derkachov:2018rot} in the particular case $D=2$. Below we show how their diagonalisation provides the S-matrix for the magnons for general $D$.

\subsection{Magnon dispersion relation}

Let us begin with the 1-magnon problem, that is the diagonalisation of the graph-building operator $\widehat{\Gamma}_{N =1}$. This operator commutes with dilatation and rotations. As such, its eigenvectors have the form $x^{-2\tilde{\beta}}C\left(x/|x|\right)$ where $C(y) = C^{\mu_1\dots\mu_l}y_{\mu_1}\dots y_{\mu_l}$ and $C$ is a symmetric traceless tensor of rank $l\in\mathbb{N}$. A complete basis of states is obtained by taking $\tilde{\beta} = (D-\tdelt)/2-\ii u$, with $u\in\mathbb{R}$, and choosing a complete basis of symmetric traceless tensors~\footnote{There is \textit{a priori} a freedom in choosing the real part of $\tilde{\beta}$, our choice makes the eigenvectors orthogonal for the pairing defined by $(f,g) = \int f^*(x_1) g(x_2) (x_1-x_2)^{-2\tdelt}\dd^D x_1\dd^D x_2$.}. One gets
\begin{equation}\label{eq:chain}
\int\frac{C\left(\frac{w}{|w|}\right)}{w^{2 \tilde{\beta}}(w-x)^{2\tdelt}}\frac{\dd^D w}{\pi^{\frac{D}{2}}}\\
= \lambda_l(u)\frac{C\left(\frac{x}{|x|}\right)}{x^{2\left(\tilde{\beta}-\delt\right)}}\, ,
\end{equation}
where the eigenvalue is given by
\begin{equation}\label{eq:dispersion}
   \lambda_l(u) = \frac{\Gamma(\delt)\Gamma\left( \frac{\tdelt}{2} + \frac{l}{2} + \ii u\right) \Gamma\left( \frac{\tdelt}{2} + \frac{l}{2} - \ii u \right)}{\Gamma(\tdelt)\Gamma\left( \frac{D-\tdelt}{2} + \frac{l}{2} + \ii u\right) \Gamma\left(\frac{D - \tdelt}{2} + \frac{l}{2} - \ii u\right)}\per
\end{equation}
This eigenvalue is the weight of propagation of a magnon with rapidity $u$ and spin $l$, it naturally defines the magnon energy $\varepsilon_l$ through $\varepsilon_l = - \log \lambda_l$ while the momentum conjugate to $\sigma$ is $p_l(u) = 2u$ as can be read off directly from the expression of the eigenvector.

\subsection{Magnon S-matrix}

\begin{figure}[!ht]
    \centering
    \includegraphics[width=0.45\textwidth]{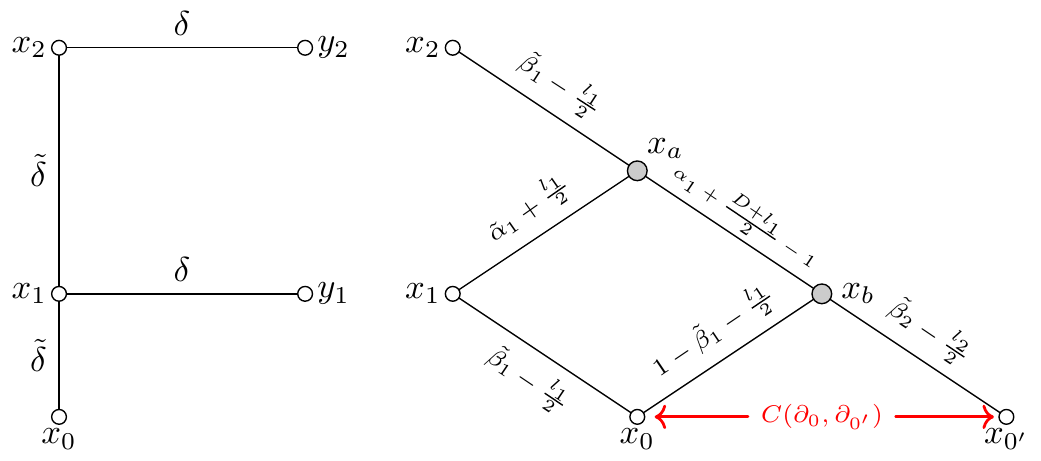}
    \caption{Integral kernel of the graph-building operator $\widehat{\Gamma}_{N=2} $ (left panel) and eigenvector \eqref{eigenvector general} (right panel). The action of the former on the latter is explained in \eqref{graph-building}. Filled blobs represent integration points while a line with index $\alpha$ connecting points $x$ and $y$ represents the propagator  $(x-y)^{-2\alpha}$.}
    \label{fig:Eigen}
\end{figure}

For the magnon S-matrix, we proceed with the diagonalization of the graph-building operator of a two-frame wheel $\widehat{\Gamma}_{N=2}$ which acts on functions of two variables. Global symmetries are no longer enough to solve the problem, but with only two magnons one can write the solution rather explicitly. We found that eigenvectors are given by
\begin{multline}
   \bra{x_1,x_2}\ket{u_1,l_1;u_2,l_2;C} = \int \frac{\dd^D x_a}{\pi^{\frac{D}{2}}} \frac{ 1}{x_{2a}^{2\left(\tilde{\beta}_1 - \frac{l_1}{2}\right)}x_{1a}^{2\left(\tilde{\alpha}_1 + \frac{l_1}{2}\right)}}\\
   C(\partial_0,\partial_{0'})\int \frac{\dd^D x_b}{\pi^{\frac{D}{2}}} \frac{x_{ab}^{-2\left(\alpha_1 + \frac{D+l_1}{2} - 1\right)}}{x_{0'b}^{2\left(\tilde{\beta}_2 - \frac{l_2}{2}\right)}} \frac{x_{0b}^{2\left(\tilde{\beta}_1 + \frac{l_1}{2} - 1\right)}}{ x_{01}^{2\left(\tilde{\beta}_1 - \frac{l_1}{2}\right)}} 
   \label{eigenvector general}
\end{multline}
evaluated at $x_0 = x_{0'} = 0$~\footnote{The integral over $x_a$ does not seem to be convergent but we believe, in view of the fact that standard manipulations allow us to prove that the full function is an eigenvector, that it should be understood as an analytic continuation.}, see figure \ref{fig:Eigen} for a graphical representation of this function. Here $x_{ij} = x_i - x_j$, $\alpha_j = D/2 -\tilde{\alpha}_j = \tdelt/2 - \ii u_j$, $\beta_j = D/2 -\tilde{\beta}_j = \tdelt/2 + \ii u_j$ and $C(y_1,y_2) = C^{\mu_1\dots\mu_{l_1}\nu_1\dots\nu_{l_2}}y_{1\mu_1}\dots y_{1\mu_{l_1}} y_{2\nu_1}\dots y_{2\nu_{l_2}}$, where $C$ is a tensor that is symmetric and traceless, separately, in the first $l_1$ indices and in the last $l_2$ ones. The eigenvalue associated to \eqref{eigenvector general} can be computed directly using notably the star-triangle identity. It reads $\lambda_{l_1}(u_1) \lambda_{l_2}(u_2)$, which means that the total energy is the sum of the individual energies.

The full S-matrix can be read off from the asymptotics of these eigenvectors in the limit $x_{12}^2\to\infty$. In this limit, the mirror magnons are far apart from each other and their wave function reduces to the sum of incoming and outgoing plane waves
\begin{multline}
    (x_1^2)^{\ii u_1}(x_2^2)^{\ii u_2} C\left(\frac{x_1}{|x_1|},\frac{x_2}{|x_2|}\right) \\
    + (x_2^2)^{\ii u_1}(x_1^2)^{\ii u_2} \left[\mathbb{S}_{l_1,l_2}(u_1,u_2)C\right] 
    \left(\frac{x_2}{|x_2|},\frac{x_1}{|x_1|}\right)
\end{multline}
up to an overall prefactor that depends on $x_1$ and $x_2$, and with, in square brackets, the tensor obtained by acting on $C$ with the S-matrix. We conjecture that the S-matrix obtained this way is given by
\be\label{eq:S-matrix}
\mathbb{S}_{l,l'}(u,v) = \frac{f_l(u)}{f_{l'}(v)} \mathcal{S}_{l, l'}(u-v) \mathbb{R}_{l,l'}(u-v)
\ee
where the dynamical factors are
\begin{align}\label{eq:S0}
\mathcal{S}_{l, l'}(u) =  
\frac{\Gamma(1+\tfrac{l+l'}{2}-\ii u)}{\Gamma(1+\tfrac{l+l'}{2}+\ii u)}
\frac{\Gamma(\frac{D}{2}+\tfrac{l+l'}{2}+\ii u)}{\Gamma(\frac{D}{2}+\tfrac{l+l'}{2}-\ii u)} \notag\\
\times\frac{\Gamma(\tfrac{|l-l'|}{2}+\ii u)}{\Gamma(\tfrac{|l-l'|}{2}-\ii u)}\frac{\Gamma(1+\tfrac{|l-l'|}{2}+\ii u)}{\Gamma(1+\tfrac{|l-l'|}{2}-\ii u)}
\end{align}
and
\be\label{eq:f}
f_l(u) = \frac{\Gamma \left(\frac{\tdelt}{2}+\frac{l}{2} -\ii u\right) \Gamma \left(\frac{D-\tdelt}{2}+\frac{l}{2}-\ii u\right)}{\Gamma \left(\frac{\tdelt}{2}+\frac{l}{2} +\ii u\right) \Gamma \left(\frac{D-\tdelt}{2}+\frac{l}{2}+\ii u\right)} \,
\ee
and where $\mathbb{R}_{l,l'}$ is the $O(D)$-symmetric rational R-matrix acting on the tensor product of spaces of symmetric traceless tensors of ranks $l$ and $l'$. These R-matrices can be determined through their eigenvalues~\cite{OGIEVETSKY1986360,Reshetikhin1983,Reshetikhin1985} or by fusing Zamolodchikov's $O(D)$ R-matrix \cite{Zamolodchikov:1978xm}
\be\label{eq:R11}
\mathbb{R}_{1,1}(u) =  \frac{u \mathbbm{1}}{u+\ii} +\frac{\ii \mathbb{P}}{u+\ii} - \frac{\ii u \mathbb{K}}{(u+\ii)(u+\ii \frac{D-2}{2})}\, ,  
\ee
with $\mathbbm{1}, \mathbb{P}$ and $\mathbb{K}$ being identity, permutation and contraction of two $D$-dimensional vectors.

We could partly verify this conjecture by considering some particular cases for which we were able to compute the asymptotic behaviour of the eigenvectors. This includes eigenvectors associated to completely symmetric traceless tensors $C$ of any rank $l_1+l_2$. They are eigenstates of $\mathbb{R}_{l_1,l_2}$ with eigenvalue $1$ and gave us access to the functions $f_l$ and $\mathcal{S}_{l_1,l_2}$. We have also explicitly reproduced $\mathbb{R}_{1,1}$. This match gives confidence that the general conjecture is correct, as the latter R-matrix is the seed for the higher-spin ones. (Indeed, assuming integrability, Yang-Baxter relations entirely determine $\mathbb{R}_{l, l'}$ given $\mathbb{R}_{1,1}$.) Nonetheless, it would be nice to check the matrix structure for higher spins from the general wave function \eqref{eigenvector general}.

For $D=2$, eigenvectors were found in \cite{Derkachov:2018rot,Derkachov:2014gya}
and we checked that their asymptotic behaviour allows one to recover exactly the full S-matrix \eqref{eq:S-matrix}-\eqref{eq:f}.

As a final check, for $D=4$ isotropic fishnets ($\delt = \tdelt$), we verify agreement with the conjectured S-matrix of the $\mathcal{N}=4$ SYM theory \cite{Beisert:2006ez} at weak coupling in the mirror kinematics. In fact, our analysis is the first field theory derivation of this mirror S-matrix. 

\section{TBA for ground state}

We turn now to the TBA equations defining the divergent part of the wheel graph on figure \ref{fig:graph}, i.e.~the contribution to the scaling dimension $\Delta$ of local operator $\mathcal{O}_{J,0}$ at a given order of perturbation theory. In the 1-dimensional picture, the scaling dimension corresponds to the free energy of a system of magnons at temperature $1/J$ and chemical potential $\log{\xi^2}$, where $J$ is the length of the operator and $\xi$ the coupling constant of the FCFT \eqref{eqn-Lag}. The factorisation of the S-matrix allows one to compute exactly the free energy at any $J$ and $\xi$. Following well known saddle-point procedure \cite{Yang:1968rm,Zamolodchikov:1989cf,Klassen:1989ui,Klassen:1990dx}, it takes the form
\be\label{eq:Energy}
\Delta = J \delt - \sum_{l \geq 0} \int p_l'(u) \log{(1+\Y_{1, l}(u))}\frac{\dd u}{2\pi}
\ee
where $p_l'(u) = 2$ and where the Y functions $\Y_{1, l}$ describe the distribution of energy per magnon, with $l \in \mathbb{N}$ labelling the spherical harmonics.

The latter Y functions are part of a larger family of functions $\lbrace\Y_{a, l}\rbrace$ needed to account for the matrix degrees of freedom. For the sake of simplicity, we shall restrict ourselves to the simply laced case, corresponding to even dimensions $D>2$. Hence, $a \in [1,r]$ labels the nodes of the $O(D+2) \cong D_{r}$ Dynkin diagram, with $r=D/2+1$ and incidence matrix $I_{ab}$.

The Y functions themselves are determined by an infinite system of non-linear TBA equations. Denoting \(\LL_{a,l}=\log(1+\Y_{a,l})\), these equations take the form 
\beq \label{eq:TBAm}
\log \Y_{1,l} = C-J \varepsilon_{l} +\sum_{l'\geq 0} \mathcal{K}_{l,l'} \star \LL_{1, l'} + \sum_{l' \geq 1} \mathrm{K}_{l,l'} \star \LL_{2, l'}\, ,
\eeq
for the massive nodes ($a=1$, $l\geq 0$), where $\varepsilon_l = -\log \lambda_l$,
\be
C = J \log \xi^2 - \sum_{l=0}^\infty \int [\ii \partial_u \log f_l(u)] \, \LL_{1,l}\frac{\dd u}{2\pi}\, ,
\ee
and where the $\star$ operation denotes the convolution on the real axis (with measure $du/2\pi$).
For the remaining, auxiliary, nodes for spin excitations ($a>1$, $l\geq 1$) the equations are
\beq \label{eq:TBAa}
\log \Y_{a,l} = \, -\sum_{l'\geq 1} \check{\mathrm{K}}_{l,l'} \star \LL_{a,l'}+ \sum_{b,I_{ab}\neq 0 }\sum_{l'\geq 1} \mathrm{K}_{l,l'} \star \LL_{b,l'}\, ,
\eeq
where we introduced $\check{\K}_{l, l'} = \K_{l, l'+1}+\K_{l, l'-1}$, with symmetric kernel $\mathcal{K},\K$ defined by
\be\label{eq:calK}
\mathcal{K}_{l,l'}(u) = -\ii \partial_u \log \mathcal{S}_{l, l'}(u) \, ,
\ee
and
\beq \label{spin-kernel}
\K_{l, l'}(u)  =\sum_{j=(|l-l'|+1)/2}^{(l+l'-1)/2}\frac{2j}{u^2+j^2}\, .
\eeq 
Finally, let us stress that the kernels obey the universal asymptotics $\mathcal{K}_{l,l'}(u) = 2\log u^2 +O(1/u^2)$ at large rapidity. Consequently, the scaling dimension \eqref{eq:Energy} controls the asymptotics of the main Y functions,
\beq\label{eq:asy}
\log{\Y_{1,l}} \sim -\Delta\log{u^2}\, .
\eeq
The auxiliary Y functions are, on the other hand, asymptotically constant at $u \rightarrow \infty$.

\section{Dual TBA and  Y-system}

The TBA equations above give us a good handle on the scaling dimension at weak coupling, which is when the massive Y functions are small and the equations are solvable iteratively. They are also very useful for the study of fishnet graphs at large order, that is when the coupling constant approaches its critical value \cite{Zamolodchikov:1980mb,Basso:2018agi}. Close to this point, the lightest ($l=0$) mirror magnons condense, driving the system towards a new phase with gapless excitations. This is analogous to the transition from ferro- to antiferromagnetic order for compact spin chains in a magnetic field. It relates to the continuum limit of the fishnet graphs and to their correspondence with $2d$ $\sigma$-models with $AdS$ target space. This correspondence, which was first discussed in~\cite{Basso:2018agi} for $D=4$, also holds in higher dimensions. Namely, there is a dual set of TBA equations looking like that of the familiar $O(D+2)$ $\sigma$-model in a finite volume $J$ except that instead of the standard relativistic dispersion relation we should use the one dual to~\eqref{eq:dispersion}.  

The duality is established by means of the familiar particle/hole transformation. It involves the operator $\mathbbm{1} - {\K}_{O(D+2)}$ which solves the equation 
\be \label{eqn-DualK}
(\mathbbm{1} - \mathcal{K}_{0,0})\star (\mathbbm{1} - \K_{O(D+2)}) = \mathbbm{1}\,, 
\ee
with $\mathbbm{1}$ the identity operator and with $\mathcal{K}_{0, 0}$ as in (\ref{eq:calK}). Straightforward algebra gives  
\be
\K_{O(D+2)}(u) = - \ii \partial_u \log S_{O(D+2)}(2\pi u/D),
\ee
with the well-known $O(D+2)$ $S$-matrix~\cite{Zamolodchikov:1978xm}:
\begin{align}
& S_{O(D+2)}(\theta) \notag\\
=& -\frac{\Gamma \left(1+\frac{\ii \theta }{2 \pi }\right) \Gamma \left(\frac{1}{2}-\frac{\ii \theta }{2 \pi }\right) \Gamma \left(\frac{1}{D}-\frac{\ii \theta }{2 \pi }\right) \Gamma \left(\frac{1}{2}+\frac{1}{D}+\frac{\ii \theta }{2 \pi }\right)}{\Gamma \left(1-\frac{\ii \theta }{2 \pi }\right) \Gamma \left(\frac{1}{2} +\frac{\ii \theta }{2 \pi }\right) \Gamma \left(\frac{1}{D} +\frac{\ii \theta }{2 \pi }\right) \Gamma \left(\frac{1}{2}+\frac{1}{D}-\frac{\ii \theta }{2 \pi }\right)}\, ,
\end{align}
hinting at the dual $\sigma$-model description.

Applying the operator $\mathbbm{1} - K_{O(D+2)}$ to \eqref{eq:TBAm} for $l=0$ we get the dual equation for the scalar node:
\beq
\log \Y_{1,0} = J E - \K_{O(D+2)} \star \LL'_{1,0} -\sum_{l\geq 1} \K_{1,l} \star \LL_{1,l},
\label{TBA dual m0}
\eeq
with $\LL'_{1,0} = \log{(1+\Y_{1,0}^{-1})}$, and the new driving term
\be
E(u) = \log \left(\frac{\cosh \left(\frac{2 \pi  u}{D}\right)+\cos \left(\frac{\pi \tdelt}{D}  \right)}{\cosh \left(\frac{2 \pi  u}{D}\right)-\cos \left(\frac{\pi \tdelt}{D}  \right)}\right)
\ee
is identified as the dual energy. As for the higher harmonics, equations \eqref{eq:TBAm} for $l>0$, they can be rewritten as
\be
\log \Y_{1,l}  = - \K_{l,1}\star \LL'_{1,0} - \sum_{l' \geq 1} \check{\K}_{l,l'}\star \LL_{1,l'} + \sum_{l'\geq 1} \K_{l,l'} \star \LL_{2,l'} \per
\label{TBA dual ml}
\ee
The absence of driving terms in these equations indicate that the full symmetry is linearly realised in the dual picture. In fact, if not for the energy, the equations \eqref{TBA dual m0} and \eqref{TBA dual ml}, as well as the ones in \eqref{eq:TBAa} which stay untouched, are identical to those for the $O(D+2)$ $\sigma$-model.

The non-compactness of the model is seen in the fact that the spectrum is gapless. This is made clearer after introducing a dual momentum $P$ obtained via a Wick rotation and a reflection $\delta \rightarrow  \tilde{\delta}$ %
\footnote{The reflection is needed because the square lattice is not invariant under a $\pi/2$ rotation when there is anisotropy.}. It yields
\be
P(u) = - \ii E\left(u + \ii D/4\right)\big|_{\delta \rightarrow  \tilde{\delta}} \, ,
\ee
resulting in the following dispersion relation
\be
\sinh^2 \frac{E}{2} = \tan^2 \left(\frac{\pi\delt}{D}\right) \times \sin^2 \frac{P}{2}\,
\ee
as for a massless particle on a square lattice. Note that the anisotropy only enters in the functional form of the energy, as expected.

In the dual TBA the fishnet coupling $\xi^2$ disappears. It is now captured by the $\sigma$-model energy $E_{2d}(\Delta, J)$, defined by
\beq
E_{2d}(\Delta,J) \equiv - \int  \LL'_{1,0} \partial_u P(u) \frac{\dd u}{2\pi} = J \log \frac{\xi^2}{\xi_{c}^2}\, ,
\eeq
with
\be 
\log \xi_c^2 = \int_0^{\infty}\left[ \frac{D}{2}e^{-t} + \frac{e^{-\tdelt t} - e^{\tdelt t}+e^{-\delt t} - e^{\delt t}} {(1-e^{-t})(1+e^{\frac{D t}{2}})} \right]\frac{\dd t}{t}
\ee 
the critical coupling, in agreement with Zamolodchikov's formula \cite{Zamolodchikov:1980mb}, up to the normalisation of the coupling constant. This match for any $\delta$ and $D$ is yet another check for our expressions.

Lastly, let us note that the TBA equations \eqref{TBA dual m0}, \eqref{TBA dual ml} and \eqref{eq:TBAa} can be brought, by inverting the kernels, to the Y-system form:
\be
\frac{\Y_{a,l}^{+}\Y_{a,l}^{-}}{\Y_{a,l+1}\Y_{a,l-1}} = \frac{\prod_{b=1}^{r} (1+\Y_{b,l})^{I_{ab}}}{(1+\Y_{a,l+1})(1+\Y_{a,l-1})},
\ee
for all nodes with \(1\le a\le r,\,\,l\ge 1\) (with the convention that $\Y_{a,0} = \infty$ for $a>1$) while $\Y_{1,0}$ satisfies
\bea
\frac{1}{\Y_{1,0}^{[r-1]}\Y_{1,0}^{[1-r]}} =
\prod_{k=1}^{r-2} &(1 + 1/\Y_{r-k-1,1}^{[k]})(1 + 1/\Y_{r-k-1,1}^{[-k]})\\
\times  &(1+ 1/\Y_{r-1,1})(1+ 1/\Y_{r,1}) \, ,
\eea
with the shorthand notation $f^{[\pm k]}(u) = f(u\pm \ii k/2)$. This agrees with the Y-system equations of the $O(2r)$ sigma model \cite{Balog:2005yz}.

\section{Excited States and Asymptotic Bethe Ansatz}

The TBA equations can be generalized to the states with an arbitrary number of magnons by the usual trick of the contour deformation \cite{Dorey:1996re,Bazhanov:1996aq}. The multi-magnon operators $\mathcal{O}_{J, M}$, associated to spiral graphs shown in right panel of figure \ref{fig:graph}, are made out of scalar magnons ($l=0$) and obtained by exciting the corresponding Y function. Most of the formulae stay the same, if not for the energy \eqref{eq:Energy} and the equations \eqref{eq:TBAm} that receive additional driving terms. In particular the anomalous dimensions $\gamma_M = \Delta-(J\tilde{\delta}+M\delta)$ of the multi-magnon states read
\beq
\gamma_{M} = \sum_{m=1}^{M}(2\ii u_{m}-\delta) - \sum_{l \geq 0} \int p_l'(u) \LL_{1, l}(u)\frac{\dd u}{2\pi}
\eeq
with the first term in the rhs coming from the logarithmic poles at $\Y_{1, 0}(u_m) = -1$. The latter conditions are the exact Bethe ansatz equations, which reduce at large $J$ and for sufficiently weak coupling to the ABA equations  
\be \label{eqn-ABA}
1 = \xi^{2J} e^{- \varepsilon_0(u_j) J} \prod_{\substack{k=1 \\ k \neq j}}^M \mathbb{S}_{0,0}(u_j, u_k)\,.
\ee
In this case, since all spins $l,l'=0$, the R matrix trivializes.
It should be supplemented with the trace cyclicity condition
\begin{equation}
    \prod_{j=1}^M \xi^{2} e^{- \varepsilon_0(u_j)} = 1\per
\end{equation}
This generalizes the $4D$ ABA equations of~\cite{Caetano:2016ydc} to any dimension $D$ and any anisotropy.

As an example, in the simplest $M=1$ case, the ABA equation predicts that the anomalous dimension is given by $(2\ii u-\delta)$,
where $u$ is the solution to
$\xi^2 e^{-\varepsilon_0(u)} =1$. Expanding $u$ perturbatively in $\xi^2$ around the classical value $-\ii\tdelt/2$, we find the one-magnon anomalous dimension
\be
\gamma_{M=1} = \frac{-2\xi^2}{\Gamma\left(\frac{D}{2}\right)} + 2 \xi^4\frac{\psi (\tdelt)+ \psi (\delt )- \psi\left(\frac{D}{2}\right)-\psi(1)}{\Gamma\left(\frac{D}{2}\right)^2} + O(\xi^6)
\ee
which agrees with the direct field theory computation.

As further checks, we considered two-magnon states for $J =5$. The corresponding mixing matrix takes the same form as for $D=4$ \cite{Caetano:2016ydc}. We have checked that the ABA prediction agrees with the direct diagrammatic computation through 3 loops. 

\section{Discussion and prospects}

We presented TBA equations for exact spectrum  of arbitrary multi-magnon operators in the fishnet CFT in any spacetime dimension $D$. These operators form an important class of local operators of the theory and contain all of the information about the mirror dynamics. There are other types of operators worth being studied, including spinning operators (i.e.~with derivatives) and the conjugate scalars $\bar Z,\bar X$. While it should be possible to include the former within the excited state TBA formalism, the latter are more elusive, and relate to the logarithmic property of the fishnet CFTs \cite{Ipsen:2018fmu,Gromov:2017cja,Caetanounpublished}. 

The most efficient form of the TBA equations is expected to be given by Baxter equations. It would be good to derive them for generic $D$ and for a general local operator. This program is already quite advanced in  $D=4$ case~\cite{Gromov:2017cja,Gromov:2019jfh,GrabnerGromovKazakovKorchemsky}, but not for other $D$'s. Our TBA equations should help filling this gap by providing important information about the analyticity conditions and so-called quantization conditions specifying the solutions. The Baxter equations formulation would also be instrumental for a thorough study of the correspondence between fishnet graphs and non-compact sigma models or to reveal relationships with string-bit models in $AdS$ \cite{Gromov:2019aku,Gromov:2019bsj,Gromov:2019jfh}. Another interesting direction concerns the generalisation of our TBAs to FCFT$_D$ supported on triangular and hexagonal fishnets, or dynamical fishnet like the one found in the context of the 3-coupling strongly twisted version of \(\mathcal{N}=4\) SYM theory~\cite{Gurdogan:2015csr,Kazakov:2018gcy}.

A natural next step in the study of FCFT$_D$ would be the computations of structure constants and multi-point correlation functions. In the mirror picture, it entails establishing the eigenfunctions of the graph-building operators in terms of Sklyanin separated variables, known so far only in $2D$~\cite{Derkachov:2001yn,Derkachov:2018rot}, see also \cite{Gromov:2019wmz} for new developments. The formalism is closely related to the hexagon approach \cite{Basso:2017jwq,Fleury:2017eph,Eden:2016xvg,Basso:2018cvy} and would permit to put it on a firmer ground for generic $D$. We believe that our two-body eigenfunction~\eqref{eigenvector general} is an important building block for Sklyanin's SoV construction for  non-compact quantum spin chains with arbitrary number of  spins in principal series representations of  $SO(1,D+1)$ symmetry~\cite{Ferrandoinprogress}, since these spin chains  have the same integrability structure as FCFT$_D$.

\textit{Note added}: While this work was in progress, we learnt from S.~Derkachev and E.~Olivucci that they had obtained the two-body wavefunction~\eqref{eigenvector general} and its $M$-body generalisation in a somewhat different form in the case of $D=4$.   

\section*{Acknowledgments}
\begin{acknowledgments}
We thank Fedor Levkovich-Maslyuk for collaboration at an early stage of this project. We are thankful to J\'anos Balog, Johannes Henn, Shota Komatsu, Yang Zhang and especially  Dmitry Chicherin, Sergey Derkachov and Enrico Olivucci for valuable comments and suggestions. We also thank Nikolay Gromov, Ivan Kostov and Konstantin Zarembo for comments on the manuscript. The work of BB and DlZ was supported by the French National Agency for Research grant ANR-17-CE31-0001-02. The work of DlZ was supported in part by the center of excellence supported by the Israel Science Foundation (grant number 2289/18). DlZ is grateful to the Max Planck Institut f\"ur Physik and CERN for the warm hospitality during the final stage of this project.

\end{acknowledgments}


\bibliography{biblio_TBA_PRL}

\begin{thebibliography}{68}%
\makeatletter
\providecommand \@ifxundefined [1]{%
 \@ifx{#1\undefined}
}%
\providecommand \@ifnum [1]{%
 \ifnum #1\expandafter \@firstoftwo
 \else \expandafter \@secondoftwo
 \fi
}%
\providecommand \@ifx [1]{%
 \ifx #1\expandafter \@firstoftwo
 \else \expandafter \@secondoftwo
 \fi
}%
\providecommand \natexlab [1]{#1}%
\providecommand \enquote  [1]{``#1''}%
\providecommand \bibnamefont  [1]{#1}%
\providecommand \bibfnamefont [1]{#1}%
\providecommand \citenamefont [1]{#1}%
\providecommand \href@noop [0]{\@secondoftwo}%
\providecommand \href [0]{\begingroup \@sanitize@url \@href}%
\providecommand \@href[1]{\@@startlink{#1}\@@href}%
\providecommand \@@href[1]{\endgroup#1\@@endlink}%
\providecommand \@sanitize@url [0]{\catcode `\\12\catcode `\$12\catcode
  `\&12\catcode `\#12\catcode `\^12\catcode `\_12\catcode `\%12\relax}%
\providecommand \@@startlink[1]{}%
\providecommand \@@endlink[0]{}%
\providecommand \url  [0]{\begingroup\@sanitize@url \@url }%
\providecommand \@url [1]{\endgroup\@href {#1}{\urlprefix }}%
\providecommand \urlprefix  [0]{URL }%
\providecommand \Eprint [0]{\href }%
\providecommand \doibase [0]{http://dx.doi.org/}%
\providecommand \selectlanguage [0]{\@gobble}%
\providecommand \bibinfo  [0]{\@secondoftwo}%
\providecommand \bibfield  [0]{\@secondoftwo}%
\providecommand \translation [1]{[#1]}%
\providecommand \BibitemOpen [0]{}%
\providecommand \bibitemStop [0]{}%
\providecommand \bibitemNoStop [0]{.\EOS\space}%
\providecommand \EOS [0]{\spacefactor3000\relax}%
\providecommand \BibitemShut  [1]{\csname bibitem#1\endcsname}%
\let\auto@bib@innerbib\@empty
\bibitem [{\citenamefont {G\"urdogan}\ and\ \citenamefont
  {Kazakov}(2016)}]{Gurdogan:2015csr}%
  \BibitemOpen
  \bibfield  {author} {\bibinfo {author} {\bibfnamefont {O.}~\bibnamefont
  {G\"urdogan}}\ and\ \bibinfo {author} {\bibfnamefont {V.}~\bibnamefont
  {Kazakov}},\ }\href {\doibase 10.1103/PhysRevLett.117.201602,
  10.1103/PhysRevLett.117.259903} {\bibfield  {journal} {\bibinfo  {journal}
  {Phys. Rev. Lett.}\ }\textbf {\bibinfo {volume} {117}},\ \bibinfo {pages}
  {201602} (\bibinfo {year} {2016})},\ \bibinfo {note} {[Addendum: Phys. Rev.
  Lett.117,no.25,259903(2016)]},\ \Eprint {http://arxiv.org/abs/1512.06704}
  {arXiv:1512.06704 [hep-th]} \BibitemShut {NoStop}%
\bibitem [{\citenamefont {Kazakov}(2018)}]{Kazakov:2018hrh}%
  \BibitemOpen
  \bibfield  {author} {\bibinfo {author} {\bibfnamefont {V.}~\bibnamefont
  {Kazakov}},\ }\href {\doibase 10.1142/9789813233867_0016,
  10.1142/S0129055X1840010X} {\ ,\ \bibinfo {pages} {293} (\bibinfo {year}
  {2018})},\ \bibinfo {note} {[Rev. Math. Phys.30,no.07,1840010(2018)]},\
  \Eprint {http://arxiv.org/abs/1802.02160} {arXiv:1802.02160 [hep-th]}
  \BibitemShut {NoStop}%
\bibitem [{\citenamefont {Gromov}\ \emph
  {et~al.}(2018{\natexlab{a}})\citenamefont {Gromov}, \citenamefont {Kazakov},\
  and\ \citenamefont {Korchemsky}}]{Gromov:2018hut}%
  \BibitemOpen
  \bibfield  {author} {\bibinfo {author} {\bibfnamefont {N.}~\bibnamefont
  {Gromov}}, \bibinfo {author} {\bibfnamefont {V.}~\bibnamefont {Kazakov}}, \
  and\ \bibinfo {author} {\bibfnamefont {G.}~\bibnamefont {Korchemsky}},\
  }\href@noop {} {\  (\bibinfo {year} {2018}{\natexlab{a}})},\ \Eprint
  {http://arxiv.org/abs/1808.02688} {arXiv:1808.02688 [hep-th]} \BibitemShut
  {NoStop}%
\bibitem [{\citenamefont {Sieg}\ and\ \citenamefont
  {Wilhelm}(2016)}]{Sieg:2016vap}%
  \BibitemOpen
  \bibfield  {author} {\bibinfo {author} {\bibfnamefont {C.}~\bibnamefont
  {Sieg}}\ and\ \bibinfo {author} {\bibfnamefont {M.}~\bibnamefont {Wilhelm}},\
  }\href {\doibase 10.1016/j.physletb.2016.03.004} {\bibfield  {journal}
  {\bibinfo  {journal} {Phys. Lett.}\ }\textbf {\bibinfo {volume} {B756}},\
  \bibinfo {pages} {118} (\bibinfo {year} {2016})},\ \Eprint
  {http://arxiv.org/abs/1602.05817} {arXiv:1602.05817 [hep-th]} \BibitemShut
  {NoStop}%
\bibitem [{\citenamefont {Karananas}\ \emph {et~al.}(2019)\citenamefont
  {Karananas}, \citenamefont {Kazakov},\ and\ \citenamefont
  {Shaposhnikov}}]{Karananas:2019fox}%
  \BibitemOpen
  \bibfield  {author} {\bibinfo {author} {\bibfnamefont {G.~K.}\ \bibnamefont
  {Karananas}}, \bibinfo {author} {\bibfnamefont {V.}~\bibnamefont {Kazakov}},
  \ and\ \bibinfo {author} {\bibfnamefont {M.}~\bibnamefont {Shaposhnikov}},\
  }\href@noop {} {\  (\bibinfo {year} {2019})},\ \Eprint
  {http://arxiv.org/abs/1908.04302} {arXiv:1908.04302 [hep-th]} \BibitemShut
  {NoStop}%
\bibitem [{\citenamefont {Caetano}\ \emph {et~al.}(2016)\citenamefont
  {Caetano}, \citenamefont {Gurdogan},\ and\ \citenamefont
  {Kazakov}}]{Caetano:2016ydc}%
  \BibitemOpen
  \bibfield  {author} {\bibinfo {author} {\bibfnamefont {J.}~\bibnamefont
  {Caetano}}, \bibinfo {author} {\bibfnamefont {O.}~\bibnamefont {Gurdogan}}, \
  and\ \bibinfo {author} {\bibfnamefont {V.}~\bibnamefont {Kazakov}},\
  }\href@noop {} {\  (\bibinfo {year} {2016})},\ \Eprint
  {http://arxiv.org/abs/1612.05895} {arXiv:1612.05895 [hep-th]} \BibitemShut
  {NoStop}%
\bibitem [{\citenamefont {Mamroud}\ and\ \citenamefont
  {Torrents}(2017)}]{Mamroud:2017uyz}%
  \BibitemOpen
  \bibfield  {author} {\bibinfo {author} {\bibfnamefont {O.}~\bibnamefont
  {Mamroud}}\ and\ \bibinfo {author} {\bibfnamefont {G.}~\bibnamefont
  {Torrents}},\ }\href {\doibase 10.1007/JHEP06(2017)012} {\bibfield  {journal}
  {\bibinfo  {journal} {JHEP}\ }\textbf {\bibinfo {volume} {06}},\ \bibinfo
  {pages} {012} (\bibinfo {year} {2017})},\ \Eprint
  {http://arxiv.org/abs/1703.04152} {arXiv:1703.04152 [hep-th]} \BibitemShut
  {NoStop}%
\bibitem [{\citenamefont {Chicherin}\ \emph {et~al.}(2018)\citenamefont
  {Chicherin}, \citenamefont {Kazakov}, \citenamefont {Loebbert}, \citenamefont
  {M{\"u}ller},\ and\ \citenamefont {Zhong}}]{Chicherin:2017cns}%
  \BibitemOpen
  \bibfield  {author} {\bibinfo {author} {\bibfnamefont {D.}~\bibnamefont
  {Chicherin}}, \bibinfo {author} {\bibfnamefont {V.}~\bibnamefont {Kazakov}},
  \bibinfo {author} {\bibfnamefont {F.}~\bibnamefont {Loebbert}}, \bibinfo
  {author} {\bibfnamefont {D.}~\bibnamefont {M{\"u}ller}}, \ and\ \bibinfo
  {author} {\bibfnamefont {D.-l.}\ \bibnamefont {Zhong}},\ }\href {\doibase
  10.1007/JHEP05(2018)003} {\bibfield  {journal} {\bibinfo  {journal} {JHEP}\
  }\textbf {\bibinfo {volume} {05}},\ \bibinfo {pages} {003} (\bibinfo {year}
  {2018})},\ \Eprint {http://arxiv.org/abs/1704.01967} {arXiv:1704.01967
  [hep-th]} \BibitemShut {NoStop}%
\bibitem [{\citenamefont {Chicherin}\ \emph {et~al.}(2017)\citenamefont
  {Chicherin}, \citenamefont {Kazakov}, \citenamefont {Loebbert}, \citenamefont
  {M{\"u}ller},\ and\ \citenamefont {Zhong}}]{Chicherin:2017frs}%
  \BibitemOpen
  \bibfield  {author} {\bibinfo {author} {\bibfnamefont {D.}~\bibnamefont
  {Chicherin}}, \bibinfo {author} {\bibfnamefont {V.}~\bibnamefont {Kazakov}},
  \bibinfo {author} {\bibfnamefont {F.}~\bibnamefont {Loebbert}}, \bibinfo
  {author} {\bibfnamefont {D.}~\bibnamefont {M{\"u}ller}}, \ and\ \bibinfo
  {author} {\bibfnamefont {D.-l.}\ \bibnamefont {Zhong}},\ }\href {\doibase
  10.1103/PhysRevD.96.121901} {\bibfield  {journal} {\bibinfo  {journal} {Phys.
  Rev.}\ }\textbf {\bibinfo {volume} {D96}},\ \bibinfo {pages} {121901}
  (\bibinfo {year} {2017})},\ \Eprint {http://arxiv.org/abs/1708.00007}
  {arXiv:1708.00007 [hep-th]} \BibitemShut {NoStop}%
\bibitem [{\citenamefont {Basso}\ and\ \citenamefont
  {Dixon}(2017)}]{Basso:2017jwq}%
  \BibitemOpen
  \bibfield  {author} {\bibinfo {author} {\bibfnamefont {B.}~\bibnamefont
  {Basso}}\ and\ \bibinfo {author} {\bibfnamefont {L.~J.}\ \bibnamefont
  {Dixon}},\ }\href {\doibase 10.1103/PhysRevLett.119.071601} {\bibfield
  {journal} {\bibinfo  {journal} {Phys. Rev. Lett.}\ }\textbf {\bibinfo
  {volume} {119}},\ \bibinfo {pages} {071601} (\bibinfo {year} {2017})},\
  \Eprint {http://arxiv.org/abs/1705.03545} {arXiv:1705.03545 [hep-th]}
  \BibitemShut {NoStop}%
\bibitem [{\citenamefont {Grabner}\ \emph {et~al.}(2018)\citenamefont
  {Grabner}, \citenamefont {Gromov}, \citenamefont {Kazakov},\ and\
  \citenamefont {Korchemsky}}]{Grabner:2017pgm}%
  \BibitemOpen
  \bibfield  {author} {\bibinfo {author} {\bibfnamefont {D.}~\bibnamefont
  {Grabner}}, \bibinfo {author} {\bibfnamefont {N.}~\bibnamefont {Gromov}},
  \bibinfo {author} {\bibfnamefont {V.}~\bibnamefont {Kazakov}}, \ and\
  \bibinfo {author} {\bibfnamefont {G.}~\bibnamefont {Korchemsky}},\ }\href
  {\doibase 10.1103/PhysRevLett.120.111601} {\bibfield  {journal} {\bibinfo
  {journal} {Phys. Rev. Lett.}\ }\textbf {\bibinfo {volume} {120}},\ \bibinfo
  {pages} {111601} (\bibinfo {year} {2018})},\ \Eprint
  {http://arxiv.org/abs/1711.04786} {arXiv:1711.04786 [hep-th]} \BibitemShut
  {NoStop}%
\bibitem [{\citenamefont {Gromov}\ and\ \citenamefont
  {Sever}(2019{\natexlab{a}})}]{Gromov:2019aku}%
  \BibitemOpen
  \bibfield  {author} {\bibinfo {author} {\bibfnamefont {N.}~\bibnamefont
  {Gromov}}\ and\ \bibinfo {author} {\bibfnamefont {A.}~\bibnamefont {Sever}},\
  }\href {\doibase 10.1103/PhysRevLett.123.081602} {\bibfield  {journal}
  {\bibinfo  {journal} {Phys. Rev. Lett.}\ }\textbf {\bibinfo {volume} {123}},\
  \bibinfo {pages} {081602} (\bibinfo {year} {2019}{\natexlab{a}})},\ \Eprint
  {http://arxiv.org/abs/1903.10508} {arXiv:1903.10508 [hep-th]} \BibitemShut
  {NoStop}%
\bibitem [{\citenamefont {Gromov}\ and\ \citenamefont
  {Sever}(2019{\natexlab{b}})}]{Gromov:2019bsj}%
  \BibitemOpen
  \bibfield  {author} {\bibinfo {author} {\bibfnamefont {N.}~\bibnamefont
  {Gromov}}\ and\ \bibinfo {author} {\bibfnamefont {A.}~\bibnamefont {Sever}},\
  }\href {\doibase 10.1007/JHEP10(2019)085} {\bibfield  {journal} {\bibinfo
  {journal} {JHEP}\ }\textbf {\bibinfo {volume} {10}},\ \bibinfo {pages} {085}
  (\bibinfo {year} {2019}{\natexlab{b}})},\ \Eprint
  {http://arxiv.org/abs/1907.01001} {arXiv:1907.01001 [hep-th]} \BibitemShut
  {NoStop}%
\bibitem [{\citenamefont {Gromov}\ and\ \citenamefont
  {Sever}(2019{\natexlab{c}})}]{Gromov:2019jfh}%
  \BibitemOpen
  \bibfield  {author} {\bibinfo {author} {\bibfnamefont {N.}~\bibnamefont
  {Gromov}}\ and\ \bibinfo {author} {\bibfnamefont {A.}~\bibnamefont {Sever}},\
  }\href@noop {} {\  (\bibinfo {year} {2019}{\natexlab{c}})},\ \Eprint
  {http://arxiv.org/abs/1908.10379} {arXiv:1908.10379 [hep-th]} \BibitemShut
  {NoStop}%
\bibitem [{\citenamefont {Gromov}\ \emph
  {et~al.}(2018{\natexlab{b}})\citenamefont {Gromov}, \citenamefont {Kazakov},
  \citenamefont {Korchemsky}, \citenamefont {Negro},\ and\ \citenamefont
  {Sizov}}]{Gromov:2017cja}%
  \BibitemOpen
  \bibfield  {author} {\bibinfo {author} {\bibfnamefont {N.}~\bibnamefont
  {Gromov}}, \bibinfo {author} {\bibfnamefont {V.}~\bibnamefont {Kazakov}},
  \bibinfo {author} {\bibfnamefont {G.}~\bibnamefont {Korchemsky}}, \bibinfo
  {author} {\bibfnamefont {S.}~\bibnamefont {Negro}}, \ and\ \bibinfo {author}
  {\bibfnamefont {G.}~\bibnamefont {Sizov}},\ }\href {\doibase
  10.1007/JHEP01(2018)095} {\bibfield  {journal} {\bibinfo  {journal} {JHEP}\
  }\textbf {\bibinfo {volume} {01}},\ \bibinfo {pages} {095} (\bibinfo {year}
  {2018}{\natexlab{b}})},\ \Eprint {http://arxiv.org/abs/1706.04167}
  {arXiv:1706.04167 [hep-th]} \BibitemShut {NoStop}%
\bibitem [{\citenamefont {Korchemsky}(2019)}]{Korchemsky:2018hnb}%
  \BibitemOpen
  \bibfield  {author} {\bibinfo {author} {\bibfnamefont {G.~P.}\ \bibnamefont
  {Korchemsky}},\ }\href {\doibase 10.1007/JHEP08(2019)028} {\bibfield
  {journal} {\bibinfo  {journal} {JHEP}\ }\textbf {\bibinfo {volume} {08}},\
  \bibinfo {pages} {028} (\bibinfo {year} {2019})},\ \Eprint
  {http://arxiv.org/abs/1812.06997} {arXiv:1812.06997 [hep-th]} \BibitemShut
  {NoStop}%
\bibitem [{\citenamefont {Kazakov}\ \emph {et~al.}(2019)\citenamefont
  {Kazakov}, \citenamefont {Olivucci},\ and\ \citenamefont
  {Preti}}]{Kazakov:2018gcy}%
  \BibitemOpen
  \bibfield  {author} {\bibinfo {author} {\bibfnamefont {V.}~\bibnamefont
  {Kazakov}}, \bibinfo {author} {\bibfnamefont {E.}~\bibnamefont {Olivucci}}, \
  and\ \bibinfo {author} {\bibfnamefont {M.}~\bibnamefont {Preti}},\ }\href
  {\doibase 10.1007/JHEP06(2019)078} {\bibfield  {journal} {\bibinfo  {journal}
  {JHEP}\ }\textbf {\bibinfo {volume} {06}},\ \bibinfo {pages} {078} (\bibinfo
  {year} {2019})},\ \Eprint {http://arxiv.org/abs/1901.00011} {arXiv:1901.00011
  [hep-th]} \BibitemShut {NoStop}%
\bibitem [{\citenamefont {Ipsen}\ \emph {et~al.}(2019)\citenamefont {Ipsen},
  \citenamefont {Staudacher},\ and\ \citenamefont {Zippelius}}]{Ipsen:2018fmu}%
  \BibitemOpen
  \bibfield  {author} {\bibinfo {author} {\bibfnamefont {A.~C.}\ \bibnamefont
  {Ipsen}}, \bibinfo {author} {\bibfnamefont {M.}~\bibnamefont {Staudacher}}, \
  and\ \bibinfo {author} {\bibfnamefont {L.}~\bibnamefont {Zippelius}},\ }\href
  {\doibase 10.1007/JHEP04(2019)044} {\bibfield  {journal} {\bibinfo  {journal}
  {JHEP}\ }\textbf {\bibinfo {volume} {04}},\ \bibinfo {pages} {044} (\bibinfo
  {year} {2019})},\ \Eprint {http://arxiv.org/abs/1812.08794} {arXiv:1812.08794
  [hep-th]} \BibitemShut {NoStop}%
\bibitem [{\citenamefont {de~Mello~Koch}\ \emph {et~al.}(2019)\citenamefont
  {de~Mello~Koch}, \citenamefont {LiMing}, \citenamefont {Van~Zyl},\ and\
  \citenamefont {Rodrigues}}]{deMelloKoch:2019ywq}%
  \BibitemOpen
  \bibfield  {author} {\bibinfo {author} {\bibfnamefont {R.}~\bibnamefont
  {de~Mello~Koch}}, \bibinfo {author} {\bibfnamefont {W.}~\bibnamefont
  {LiMing}}, \bibinfo {author} {\bibfnamefont {H.~J.~R.}\ \bibnamefont
  {Van~Zyl}}, \ and\ \bibinfo {author} {\bibfnamefont {J.~P.}\ \bibnamefont
  {Rodrigues}},\ }\href {\doibase 10.1016/j.physletb.2019.04.044} {\bibfield
  {journal} {\bibinfo  {journal} {Phys. Lett.}\ }\textbf {\bibinfo {volume}
  {B793}},\ \bibinfo {pages} {169} (\bibinfo {year} {2019})},\ \Eprint
  {http://arxiv.org/abs/1902.06409} {arXiv:1902.06409 [hep-th]} \BibitemShut
  {NoStop}%
\bibitem [{\citenamefont {Pittelli}\ and\ \citenamefont
  {Preti}(2019)}]{Pittelli:2019ceq}%
  \BibitemOpen
  \bibfield  {author} {\bibinfo {author} {\bibfnamefont {A.}~\bibnamefont
  {Pittelli}}\ and\ \bibinfo {author} {\bibfnamefont {M.}~\bibnamefont
  {Preti}},\ }\href {\doibase 10.1016/j.physletb.2019.134971} {\  (\bibinfo
  {year} {2019}),\ 10.1016/j.physletb.2019.134971},\ \Eprint
  {http://arxiv.org/abs/1906.03680} {arXiv:1906.03680 [hep-th]} \BibitemShut
  {NoStop}%
\bibitem [{\citenamefont {Dutta~Chowdhury}\ \emph {et~al.}(2019)\citenamefont
  {Dutta~Chowdhury}, \citenamefont {Haldar},\ and\ \citenamefont
  {Sen}}]{Chowdhury:2019hns}%
  \BibitemOpen
  \bibfield  {author} {\bibinfo {author} {\bibfnamefont {S.}~\bibnamefont
  {Dutta~Chowdhury}}, \bibinfo {author} {\bibfnamefont {P.}~\bibnamefont
  {Haldar}}, \ and\ \bibinfo {author} {\bibfnamefont {K.}~\bibnamefont {Sen}},\
  }\href {\doibase 10.1007/JHEP10(2019)249} {\bibfield  {journal} {\bibinfo
  {journal} {JHEP}\ }\textbf {\bibinfo {volume} {10}},\ \bibinfo {pages} {249}
  (\bibinfo {year} {2019})},\ \Eprint {http://arxiv.org/abs/1908.01123}
  {arXiv:1908.01123 [hep-th]} \BibitemShut {NoStop}%
\bibitem [{\citenamefont {Adamo}\ and\ \citenamefont
  {Jaitly}(2019)}]{Adamo:2019lor}%
  \BibitemOpen
  \bibfield  {author} {\bibinfo {author} {\bibfnamefont {T.}~\bibnamefont
  {Adamo}}\ and\ \bibinfo {author} {\bibfnamefont {S.}~\bibnamefont {Jaitly}},\
  }\href@noop {} {\  (\bibinfo {year} {2019})},\ \Eprint
  {http://arxiv.org/abs/1908.11220} {arXiv:1908.11220 [hep-th]} \BibitemShut
  {NoStop}%
\bibitem [{\citenamefont {Kazakov}\ and\ \citenamefont
  {Olivucci}(2018)}]{Kazakov:2018qbr}%
  \BibitemOpen
  \bibfield  {author} {\bibinfo {author} {\bibfnamefont {V.}~\bibnamefont
  {Kazakov}}\ and\ \bibinfo {author} {\bibfnamefont {E.}~\bibnamefont
  {Olivucci}},\ }\href {\doibase 10.1103/PhysRevLett.121.131601} {\bibfield
  {journal} {\bibinfo  {journal} {Phys. Rev. Lett.}\ }\textbf {\bibinfo
  {volume} {121}},\ \bibinfo {pages} {131601} (\bibinfo {year} {2018})},\
  \Eprint {http://arxiv.org/abs/1801.09844} {arXiv:1801.09844 [hep-th]}
  \BibitemShut {NoStop}%
\bibitem [{\citenamefont {Zamolodchikov}(1980)}]{Zamolodchikov:1980mb}%
  \BibitemOpen
  \bibfield  {author} {\bibinfo {author} {\bibfnamefont {A.~B.}\ \bibnamefont
  {Zamolodchikov}},\ }\href {\doibase 10.1016/0370-2693(80)90547-X} {\bibfield
  {journal} {\bibinfo  {journal} {Phys. Lett.}\ }\textbf {\bibinfo {volume}
  {97B}},\ \bibinfo {pages} {63} (\bibinfo {year} {1980})}\BibitemShut
  {NoStop}%
\bibitem [{Note1()}]{Note1}%
  \BibitemOpen
  \bibinfo {note} {The structure of such graph near its boundary depends on the
  computed physical quantity}\BibitemShut {NoStop}%
\bibitem [{\citenamefont {Chicherin}\ \emph {et~al.}(2013)\citenamefont
  {Chicherin}, \citenamefont {Derkachov},\ and\ \citenamefont
  {Isaev}}]{Chicherin:2012yn}%
  \BibitemOpen
  \bibfield  {author} {\bibinfo {author} {\bibfnamefont {D.}~\bibnamefont
  {Chicherin}}, \bibinfo {author} {\bibfnamefont {S.}~\bibnamefont
  {Derkachov}}, \ and\ \bibinfo {author} {\bibfnamefont {A.~P.}\ \bibnamefont
  {Isaev}},\ }\href {\doibase 10.1007/JHEP04(2013)020} {\bibfield  {journal}
  {\bibinfo  {journal} {JHEP}\ }\textbf {\bibinfo {volume} {04}},\ \bibinfo
  {pages} {020} (\bibinfo {year} {2013})},\ \Eprint
  {http://arxiv.org/abs/1206.4150} {arXiv:1206.4150 [math-ph]} \BibitemShut
  {NoStop}%
\bibitem [{\citenamefont {Beisert}\ \emph {et~al.}(2012)\citenamefont {Beisert}
  \emph {et~al.}}]{Beisert:2010jr}%
  \BibitemOpen
  \bibfield  {author} {\bibinfo {author} {\bibfnamefont {N.}~\bibnamefont
  {Beisert}} \emph {et~al.},\ }\href {\doibase 10.1007/s11005-011-0529-2}
  {\bibfield  {journal} {\bibinfo  {journal} {Lett. Math. Phys.}\ }\textbf
  {\bibinfo {volume} {99}},\ \bibinfo {pages} {3} (\bibinfo {year} {2012})},\
  \Eprint {http://arxiv.org/abs/1012.3982} {arXiv:1012.3982 [hep-th]}
  \BibitemShut {NoStop}%
\bibitem [{\citenamefont {Beisert}\ and\ \citenamefont
  {Staudacher}(2005)}]{Beisert:2005fw}%
  \BibitemOpen
  \bibfield  {author} {\bibinfo {author} {\bibfnamefont {N.}~\bibnamefont
  {Beisert}}\ and\ \bibinfo {author} {\bibfnamefont {M.}~\bibnamefont
  {Staudacher}},\ }\href {\doibase 10.1016/j.nuclphysb.2005.06.038} {\bibfield
  {journal} {\bibinfo  {journal} {Nucl. Phys.}\ }\textbf {\bibinfo {volume}
  {B727}},\ \bibinfo {pages} {1} (\bibinfo {year} {2005})},\ \Eprint
  {http://arxiv.org/abs/hep-th/0504190} {arXiv:hep-th/0504190 [hep-th]}
  \BibitemShut {NoStop}%
\bibitem [{\citenamefont {Beisert}\ \emph {et~al.}(2007)\citenamefont
  {Beisert}, \citenamefont {Eden},\ and\ \citenamefont
  {Staudacher}}]{Beisert:2006ez}%
  \BibitemOpen
  \bibfield  {author} {\bibinfo {author} {\bibfnamefont {N.}~\bibnamefont
  {Beisert}}, \bibinfo {author} {\bibfnamefont {B.}~\bibnamefont {Eden}}, \
  and\ \bibinfo {author} {\bibfnamefont {M.}~\bibnamefont {Staudacher}},\
  }\href {\doibase 10.1088/1742-5468/2007/01/P01021} {\bibfield  {journal}
  {\bibinfo  {journal} {J. Stat. Mech.}\ }\textbf {\bibinfo {volume} {0701}},\
  \bibinfo {pages} {P01021} (\bibinfo {year} {2007})},\ \Eprint
  {http://arxiv.org/abs/hep-th/0610251} {arXiv:hep-th/0610251 [hep-th]}
  \BibitemShut {NoStop}%
\bibitem [{\citenamefont {Gromov}\ \emph {et~al.}(2009)\citenamefont {Gromov},
  \citenamefont {Kazakov},\ and\ \citenamefont {Vieira}}]{Gromov:2009tv}%
  \BibitemOpen
  \bibfield  {author} {\bibinfo {author} {\bibfnamefont {N.}~\bibnamefont
  {Gromov}}, \bibinfo {author} {\bibfnamefont {V.}~\bibnamefont {Kazakov}}, \
  and\ \bibinfo {author} {\bibfnamefont {P.}~\bibnamefont {Vieira}},\ }\href
  {\doibase 10.1103/PhysRevLett.103.131601} {\bibfield  {journal} {\bibinfo
  {journal} {Phys. Rev. Lett.}\ }\textbf {\bibinfo {volume} {103}},\ \bibinfo
  {pages} {131601} (\bibinfo {year} {2009})},\ \Eprint
  {http://arxiv.org/abs/0901.3753} {arXiv:0901.3753 [hep-th]} \BibitemShut
  {NoStop}%
\bibitem [{\citenamefont {Basso}\ \emph {et~al.}(2015)\citenamefont {Basso},
  \citenamefont {Komatsu},\ and\ \citenamefont {Vieira}}]{Basso:2015zoa}%
  \BibitemOpen
  \bibfield  {author} {\bibinfo {author} {\bibfnamefont {B.}~\bibnamefont
  {Basso}}, \bibinfo {author} {\bibfnamefont {S.}~\bibnamefont {Komatsu}}, \
  and\ \bibinfo {author} {\bibfnamefont {P.}~\bibnamefont {Vieira}},\
  }\href@noop {} {\  (\bibinfo {year} {2015})},\ \Eprint
  {http://arxiv.org/abs/1505.06745} {arXiv:1505.06745 [hep-th]} \BibitemShut
  {NoStop}%
\bibitem [{\citenamefont {Fleury}\ and\ \citenamefont
  {Komatsu}(2018)}]{Fleury:2017eph}%
  \BibitemOpen
  \bibfield  {author} {\bibinfo {author} {\bibfnamefont {T.}~\bibnamefont
  {Fleury}}\ and\ \bibinfo {author} {\bibfnamefont {S.}~\bibnamefont
  {Komatsu}},\ }\href {\doibase 10.1007/JHEP02(2018)177} {\bibfield  {journal}
  {\bibinfo  {journal} {JHEP}\ }\textbf {\bibinfo {volume} {02}},\ \bibinfo
  {pages} {177} (\bibinfo {year} {2018})},\ \Eprint
  {http://arxiv.org/abs/1711.05327} {arXiv:1711.05327 [hep-th]} \BibitemShut
  {NoStop}%
\bibitem [{\citenamefont {Eden}\ and\ \citenamefont
  {Sfondrini}(2017)}]{Eden:2016xvg}%
  \BibitemOpen
  \bibfield  {author} {\bibinfo {author} {\bibfnamefont {B.}~\bibnamefont
  {Eden}}\ and\ \bibinfo {author} {\bibfnamefont {A.}~\bibnamefont
  {Sfondrini}},\ }\href {\doibase 10.1007/JHEP10(2017)098} {\bibfield
  {journal} {\bibinfo  {journal} {JHEP}\ }\textbf {\bibinfo {volume} {10}},\
  \bibinfo {pages} {098} (\bibinfo {year} {2017})},\ \Eprint
  {http://arxiv.org/abs/1611.05436} {arXiv:1611.05436 [hep-th]} \BibitemShut
  {NoStop}%
\bibitem [{\citenamefont {Bombardelli}\ \emph {et~al.}(2009)\citenamefont
  {Bombardelli}, \citenamefont {Fioravanti},\ and\ \citenamefont
  {Tateo}}]{Bombardelli:2009ns}%
  \BibitemOpen
  \bibfield  {author} {\bibinfo {author} {\bibfnamefont {D.}~\bibnamefont
  {Bombardelli}}, \bibinfo {author} {\bibfnamefont {D.}~\bibnamefont
  {Fioravanti}}, \ and\ \bibinfo {author} {\bibfnamefont {R.}~\bibnamefont
  {Tateo}},\ }\href {\doibase 10.1088/1751-8113/42/37/375401} {\bibfield
  {journal} {\bibinfo  {journal} {J. Phys.}\ }\textbf {\bibinfo {volume}
  {A42}},\ \bibinfo {pages} {375401} (\bibinfo {year} {2009})},\ \Eprint
  {http://arxiv.org/abs/0902.3930} {arXiv:0902.3930 [hep-th]} \BibitemShut
  {NoStop}%
\bibitem [{\citenamefont {Gromov}\ \emph {et~al.}(2010)\citenamefont {Gromov},
  \citenamefont {Kazakov}, \citenamefont {Kozak},\ and\ \citenamefont
  {Vieira}}]{Gromov:2009bc}%
  \BibitemOpen
  \bibfield  {author} {\bibinfo {author} {\bibfnamefont {N.}~\bibnamefont
  {Gromov}}, \bibinfo {author} {\bibfnamefont {V.}~\bibnamefont {Kazakov}},
  \bibinfo {author} {\bibfnamefont {A.}~\bibnamefont {Kozak}}, \ and\ \bibinfo
  {author} {\bibfnamefont {P.}~\bibnamefont {Vieira}},\ }\href {\doibase
  10.1007/s11005-010-0374-8} {\bibfield  {journal} {\bibinfo  {journal} {Lett.
  Math. Phys.}\ }\textbf {\bibinfo {volume} {91}},\ \bibinfo {pages} {265}
  (\bibinfo {year} {2010})},\ \Eprint {http://arxiv.org/abs/0902.4458}
  {arXiv:0902.4458 [hep-th]} \BibitemShut {NoStop}%
\bibitem [{\citenamefont {Arutyunov}\ and\ \citenamefont
  {Frolov}(2009)}]{Arutyunov:2009ur}%
  \BibitemOpen
  \bibfield  {author} {\bibinfo {author} {\bibfnamefont {G.}~\bibnamefont
  {Arutyunov}}\ and\ \bibinfo {author} {\bibfnamefont {S.}~\bibnamefont
  {Frolov}},\ }\href {\doibase 10.1088/1126-6708/2009/05/068} {\bibfield
  {journal} {\bibinfo  {journal} {JHEP}\ }\textbf {\bibinfo {volume} {05}},\
  \bibinfo {pages} {068} (\bibinfo {year} {2009})},\ \Eprint
  {http://arxiv.org/abs/0903.0141} {arXiv:0903.0141 [hep-th]} \BibitemShut
  {NoStop}%
\bibitem [{\citenamefont {Gromov}\ \emph {et~al.}(2014)\citenamefont {Gromov},
  \citenamefont {Kazakov}, \citenamefont {Leurent},\ and\ \citenamefont
  {Volin}}]{Gromov:2013pga}%
  \BibitemOpen
  \bibfield  {author} {\bibinfo {author} {\bibfnamefont {N.}~\bibnamefont
  {Gromov}}, \bibinfo {author} {\bibfnamefont {V.}~\bibnamefont {Kazakov}},
  \bibinfo {author} {\bibfnamefont {S.}~\bibnamefont {Leurent}}, \ and\
  \bibinfo {author} {\bibfnamefont {D.}~\bibnamefont {Volin}},\ }\href
  {\doibase 10.1103/PhysRevLett.112.011602} {\bibfield  {journal} {\bibinfo
  {journal} {Phys. Rev. Lett.}\ }\textbf {\bibinfo {volume} {112}},\ \bibinfo
  {pages} {011602} (\bibinfo {year} {2014})},\ \Eprint
  {http://arxiv.org/abs/1305.1939} {arXiv:1305.1939 [hep-th]} \BibitemShut
  {NoStop}%
\bibitem [{\citenamefont {Gromov}\ \emph {et~al.}(2015)\citenamefont {Gromov},
  \citenamefont {Kazakov}, \citenamefont {Leurent},\ and\ \citenamefont
  {Volin}}]{Gromov:2014caa}%
  \BibitemOpen
  \bibfield  {author} {\bibinfo {author} {\bibfnamefont {N.}~\bibnamefont
  {Gromov}}, \bibinfo {author} {\bibfnamefont {V.}~\bibnamefont {Kazakov}},
  \bibinfo {author} {\bibfnamefont {S.}~\bibnamefont {Leurent}}, \ and\
  \bibinfo {author} {\bibfnamefont {D.}~\bibnamefont {Volin}},\ }\href
  {\doibase 10.1007/JHEP09(2015)187} {\bibfield  {journal} {\bibinfo  {journal}
  {JHEP}\ }\textbf {\bibinfo {volume} {09}},\ \bibinfo {pages} {187} (\bibinfo
  {year} {2015})},\ \Eprint {http://arxiv.org/abs/1405.4857} {arXiv:1405.4857
  [hep-th]} \BibitemShut {NoStop}%
\bibitem [{\citenamefont {Panzer}(2015)}]{Panzer:2015ida}%
  \BibitemOpen
  \bibfield  {author} {\bibinfo {author} {\bibfnamefont {E.}~\bibnamefont
  {Panzer}},\ }\emph {\bibinfo {title} {{Feynman integrals and
  hyperlogarithms}}},\ \href
  {https://inspirehep.net/record/1377774/files/arXiv:1506.07243.pdf} {Ph.D.
  thesis},\ \bibinfo  {school} {Humboldt U., Berlin, Inst. Math.} (\bibinfo
  {year} {2015}),\ \Eprint {http://arxiv.org/abs/1506.07243} {arXiv:1506.07243
  [math-ph]} \BibitemShut {NoStop}%
\bibitem [{\citenamefont {Broadhurst}(1985)}]{Broadhurst:1985vq}%
  \BibitemOpen
  \bibfield  {author} {\bibinfo {author} {\bibfnamefont {D.~J.}\ \bibnamefont
  {Broadhurst}},\ }\href {\doibase 10.1016/0370-2693(85)90340-5} {\bibfield
  {journal} {\bibinfo  {journal} {Phys. Lett.}\ }\textbf {\bibinfo {volume}
  {B164}},\ \bibinfo {pages} {356} (\bibinfo {year} {1985})}\BibitemShut
  {NoStop}%
\bibitem [{\citenamefont {Derkachov}\ \emph {et~al.}(2019)\citenamefont
  {Derkachov}, \citenamefont {Kazakov},\ and\ \citenamefont
  {Olivucci}}]{Derkachov:2018rot}%
  \BibitemOpen
  \bibfield  {author} {\bibinfo {author} {\bibfnamefont {S.}~\bibnamefont
  {Derkachov}}, \bibinfo {author} {\bibfnamefont {V.}~\bibnamefont {Kazakov}},
  \ and\ \bibinfo {author} {\bibfnamefont {E.}~\bibnamefont {Olivucci}},\
  }\href {\doibase 10.1007/JHEP04(2019)032} {\bibfield  {journal} {\bibinfo
  {journal} {JHEP}\ }\textbf {\bibinfo {volume} {04}},\ \bibinfo {pages} {032}
  (\bibinfo {year} {2019})},\ \Eprint {http://arxiv.org/abs/1811.10623}
  {arXiv:1811.10623 [hep-th]} \BibitemShut {NoStop}%
\bibitem [{\citenamefont {Ahn}\ \emph {et~al.}(2011)\citenamefont {Ahn},
  \citenamefont {Bajnok}, \citenamefont {Bombardelli},\ and\ \citenamefont
  {Nepomechie}}]{Ahn:2011xq}%
  \BibitemOpen
  \bibfield  {author} {\bibinfo {author} {\bibfnamefont {C.}~\bibnamefont
  {Ahn}}, \bibinfo {author} {\bibfnamefont {Z.}~\bibnamefont {Bajnok}},
  \bibinfo {author} {\bibfnamefont {D.}~\bibnamefont {Bombardelli}}, \ and\
  \bibinfo {author} {\bibfnamefont {R.~I.}\ \bibnamefont {Nepomechie}},\ }\href
  {\doibase 10.1007/JHEP12(2011)059} {\bibfield  {journal} {\bibinfo  {journal}
  {JHEP}\ }\textbf {\bibinfo {volume} {12}},\ \bibinfo {pages} {059} (\bibinfo
  {year} {2011})},\ \Eprint {http://arxiv.org/abs/1108.4914} {arXiv:1108.4914
  [hep-th]} \BibitemShut {NoStop}%
\bibitem [{Note2()}]{Note2}%
  \BibitemOpen
  \bibinfo {note} {For \(D=3\), the ABJM model, in a similar double scaling
  limit, becomes a FCFT dominated by regular triangular planar graphs~\cite
  {Caetano:2016ydc}}\BibitemShut {NoStop}%
\bibitem [{\citenamefont {Beisert}(2008)}]{Beisert:2005tm}%
  \BibitemOpen
  \bibfield  {author} {\bibinfo {author} {\bibfnamefont {N.}~\bibnamefont
  {Beisert}},\ }\href {\doibase 10.4310/ATMP.2008.v12.n5.a1} {\bibfield
  {journal} {\bibinfo  {journal} {Adv. Theor. Math. Phys.}\ }\textbf {\bibinfo
  {volume} {12}},\ \bibinfo {pages} {945} (\bibinfo {year} {2008})},\ \Eprint
  {http://arxiv.org/abs/hep-th/0511082} {arXiv:hep-th/0511082 [hep-th]}
  \BibitemShut {NoStop}%
\bibitem [{\citenamefont {Janik}(2006)}]{Janik:2006dc}%
  \BibitemOpen
  \bibfield  {author} {\bibinfo {author} {\bibfnamefont {R.~A.}\ \bibnamefont
  {Janik}},\ }\href {\doibase 10.1103/PhysRevD.73.086006} {\bibfield  {journal}
  {\bibinfo  {journal} {Phys. Rev.}\ }\textbf {\bibinfo {volume} {D73}},\
  \bibinfo {pages} {086006} (\bibinfo {year} {2006})},\ \Eprint
  {http://arxiv.org/abs/hep-th/0603038} {arXiv:hep-th/0603038 [hep-th]}
  \BibitemShut {NoStop}%
\bibitem [{\citenamefont {Beisert}\ \emph {et~al.}(2006)\citenamefont
  {Beisert}, \citenamefont {Hernandez},\ and\ \citenamefont
  {Lopez}}]{Beisert:2006ib}%
  \BibitemOpen
  \bibfield  {author} {\bibinfo {author} {\bibfnamefont {N.}~\bibnamefont
  {Beisert}}, \bibinfo {author} {\bibfnamefont {R.}~\bibnamefont {Hernandez}},
  \ and\ \bibinfo {author} {\bibfnamefont {E.}~\bibnamefont {Lopez}},\ }\href
  {\doibase 10.1088/1126-6708/2006/11/070} {\bibfield  {journal} {\bibinfo
  {journal} {JHEP}\ }\textbf {\bibinfo {volume} {11}},\ \bibinfo {pages} {070}
  (\bibinfo {year} {2006})},\ \Eprint {http://arxiv.org/abs/hep-th/0609044}
  {arXiv:hep-th/0609044 [hep-th]} \BibitemShut {NoStop}%
\bibitem [{Note3()}]{Note3}%
  \BibitemOpen
  \bibinfo {note} {There is \protect \textit {a priori} a freedom in choosing
  the real part of $\protect \mathaccentV {tilde}07E{\beta }$, our choice makes
  the eigenvectors orthogonal for the pairing defined by $(f,g) = \DOTSI \intop
  \ilimits@ f^*(x_1) g(x_2) (x_1-x_2)^{-2\delta }\dd ^D x_1\dd ^D
  x_2$.}\BibitemShut {Stop}%
\bibitem [{Note4()}]{Note4}%
  \BibitemOpen
  \bibinfo {note} {The integral over $x_a$ does not seem to be convergent but
  we believe, in view of the fact that standard manipulations allow us to prove
  that the full function is an eigenvector, that it should be understood as an
  analytic continuation.}\BibitemShut {Stop}%
\bibitem [{\citenamefont {Ogievetsky}\ and\ \citenamefont
  {Wiegmann}(1986)}]{OGIEVETSKY1986360}%
  \BibitemOpen
  \bibfield  {author} {\bibinfo {author} {\bibfnamefont {E.}~\bibnamefont
  {Ogievetsky}}\ and\ \bibinfo {author} {\bibfnamefont {P.}~\bibnamefont
  {Wiegmann}},\ }\href {\doibase https://doi.org/10.1016/0370-2693(86)91644-8}
  {\bibfield  {journal} {\bibinfo  {journal} {Physics Letters B}\ }\textbf
  {\bibinfo {volume} {168}},\ \bibinfo {pages} {360 } (\bibinfo {year}
  {1986})}\BibitemShut {NoStop}%
\bibitem [{\citenamefont {Reshetikhin}(1983)}]{Reshetikhin1983}%
  \BibitemOpen
  \bibfield  {author} {\bibinfo {author} {\bibfnamefont {N.~Y.}\ \bibnamefont
  {Reshetikhin}},\ }\href {\doibase 10.1007/BF00400435} {\bibfield  {journal}
  {\bibinfo  {journal} {Letters in Mathematical Physics}\ }\textbf {\bibinfo
  {volume} {7}},\ \bibinfo {pages} {205} (\bibinfo {year} {1983})}\BibitemShut
  {NoStop}%
\bibitem [{\citenamefont {Reshetikhin}(1985)}]{Reshetikhin1985}%
  \BibitemOpen
  \bibfield  {author} {\bibinfo {author} {\bibfnamefont {N.~Y.}\ \bibnamefont
  {Reshetikhin}},\ }\href {\doibase 10.1007/BF01017501} {\bibfield  {journal}
  {\bibinfo  {journal} {Theoretical and Mathematical Physics}\ }\textbf
  {\bibinfo {volume} {63}},\ \bibinfo {pages} {555} (\bibinfo {year}
  {1985})}\BibitemShut {NoStop}%
\bibitem [{\citenamefont {Zamolodchikov}\ and\ \citenamefont
  {Zamolodchikov}(1979)}]{Zamolodchikov:1978xm}%
  \BibitemOpen
  \bibfield  {author} {\bibinfo {author} {\bibfnamefont {A.~B.}\ \bibnamefont
  {Zamolodchikov}}\ and\ \bibinfo {author} {\bibfnamefont {A.~B.}\ \bibnamefont
  {Zamolodchikov}},\ }\href {\doibase 10.1016/0003-4916(79)90391-9} {\bibfield
  {journal} {\bibinfo  {journal} {Annals Phys.}\ }\textbf {\bibinfo {volume}
  {120}},\ \bibinfo {pages} {253} (\bibinfo {year} {1979})},\ \bibinfo {note}
  {[,559(1978)]}\BibitemShut {NoStop}%
\bibitem [{\citenamefont {Derkachov}\ and\ \citenamefont
  {Manashov}(2014)}]{Derkachov:2014gya}%
  \BibitemOpen
  \bibfield  {author} {\bibinfo {author} {\bibfnamefont {S.~E.}\ \bibnamefont
  {Derkachov}}\ and\ \bibinfo {author} {\bibfnamefont {A.~N.}\ \bibnamefont
  {Manashov}},\ }\href {\doibase 10.1088/1751-8113/47/30/305204} {\bibfield
  {journal} {\bibinfo  {journal} {J. Phys.}\ }\textbf {\bibinfo {volume}
  {A47}},\ \bibinfo {pages} {305204} (\bibinfo {year} {2014})},\ \Eprint
  {http://arxiv.org/abs/1401.7477} {arXiv:1401.7477 [math-ph]} \BibitemShut
  {NoStop}%
\bibitem [{\citenamefont {Yang}\ and\ \citenamefont
  {Yang}(1969)}]{Yang:1968rm}%
  \BibitemOpen
  \bibfield  {author} {\bibinfo {author} {\bibfnamefont {C.-N.}\ \bibnamefont
  {Yang}}\ and\ \bibinfo {author} {\bibfnamefont {C.~P.}\ \bibnamefont
  {Yang}},\ }\href {\doibase 10.1063/1.1664947} {\bibfield  {journal} {\bibinfo
   {journal} {J. Math. Phys.}\ }\textbf {\bibinfo {volume} {10}},\ \bibinfo
  {pages} {1115} (\bibinfo {year} {1969})},\ \bibinfo {note}
  {[,410(1968)]}\BibitemShut {NoStop}%
\bibitem [{\citenamefont {Zamolodchikov}(1990)}]{Zamolodchikov:1989cf}%
  \BibitemOpen
  \bibfield  {author} {\bibinfo {author} {\bibfnamefont {A.~B.}\ \bibnamefont
  {Zamolodchikov}},\ }\href {\doibase 10.1016/0550-3213(90)90333-9} {\bibfield
  {journal} {\bibinfo  {journal} {Nucl. Phys.}\ }\textbf {\bibinfo {volume}
  {B342}},\ \bibinfo {pages} {695} (\bibinfo {year} {1990})}\BibitemShut
  {NoStop}%
\bibitem [{\citenamefont {Klassen}\ and\ \citenamefont
  {Melzer}(1990)}]{Klassen:1989ui}%
  \BibitemOpen
  \bibfield  {author} {\bibinfo {author} {\bibfnamefont {T.~R.}\ \bibnamefont
  {Klassen}}\ and\ \bibinfo {author} {\bibfnamefont {E.}~\bibnamefont
  {Melzer}},\ }\href {\doibase 10.1016/0550-3213(90)90643-R} {\bibfield
  {journal} {\bibinfo  {journal} {Nucl. Phys.}\ }\textbf {\bibinfo {volume}
  {B338}},\ \bibinfo {pages} {485} (\bibinfo {year} {1990})}\BibitemShut
  {NoStop}%
\bibitem [{\citenamefont {Klassen}\ and\ \citenamefont
  {Melzer}(1991)}]{Klassen:1990dx}%
  \BibitemOpen
  \bibfield  {author} {\bibinfo {author} {\bibfnamefont {T.~R.}\ \bibnamefont
  {Klassen}}\ and\ \bibinfo {author} {\bibfnamefont {E.}~\bibnamefont
  {Melzer}},\ }\href {\doibase 10.1016/0550-3213(91)90159-U} {\bibfield
  {journal} {\bibinfo  {journal} {Nucl. Phys.}\ }\textbf {\bibinfo {volume}
  {B350}},\ \bibinfo {pages} {635} (\bibinfo {year} {1991})}\BibitemShut
  {NoStop}%
\bibitem [{\citenamefont {Basso}\ and\ \citenamefont
  {Zhong}(2019)}]{Basso:2018agi}%
  \BibitemOpen
  \bibfield  {author} {\bibinfo {author} {\bibfnamefont {B.}~\bibnamefont
  {Basso}}\ and\ \bibinfo {author} {\bibfnamefont {D.-l.}\ \bibnamefont
  {Zhong}},\ }\href {\doibase 10.1007/JHEP01(2019)002} {\bibfield  {journal}
  {\bibinfo  {journal} {JHEP}\ }\textbf {\bibinfo {volume} {01}},\ \bibinfo
  {pages} {002} (\bibinfo {year} {2019})},\ \Eprint
  {http://arxiv.org/abs/1806.04105} {arXiv:1806.04105 [hep-th]} \BibitemShut
  {NoStop}%
\bibitem [{Note5()}]{Note5}%
  \BibitemOpen
  \bibinfo {note} {The reflection is needed because the square lattice is not
  invariant under a $\pi /2$ rotation when there is anisotropy.}\BibitemShut
  {Stop}%
\bibitem [{\citenamefont {Balog}\ and\ \citenamefont
  {Hegedus}(2005)}]{Balog:2005yz}%
  \BibitemOpen
  \bibfield  {author} {\bibinfo {author} {\bibfnamefont {J.}~\bibnamefont
  {Balog}}\ and\ \bibinfo {author} {\bibfnamefont {A.}~\bibnamefont
  {Hegedus}},\ }\href {\doibase 10.1016/j.nuclphysb.2005.07.032} {\bibfield
  {journal} {\bibinfo  {journal} {Nucl. Phys.}\ }\textbf {\bibinfo {volume}
  {B725}},\ \bibinfo {pages} {531} (\bibinfo {year} {2005})},\ \Eprint
  {http://arxiv.org/abs/hep-th/0504186} {arXiv:hep-th/0504186 [hep-th]}
  \BibitemShut {NoStop}%
\bibitem [{\citenamefont {Dorey}\ and\ \citenamefont
  {Tateo}(1996)}]{Dorey:1996re}%
  \BibitemOpen
  \bibfield  {author} {\bibinfo {author} {\bibfnamefont {P.}~\bibnamefont
  {Dorey}}\ and\ \bibinfo {author} {\bibfnamefont {R.}~\bibnamefont {Tateo}},\
  }\href {\doibase 10.1016/S0550-3213(96)00516-0} {\bibfield  {journal}
  {\bibinfo  {journal} {Nucl. Phys.}\ }\textbf {\bibinfo {volume} {B482}},\
  \bibinfo {pages} {639} (\bibinfo {year} {1996})},\ \Eprint
  {http://arxiv.org/abs/hep-th/9607167} {arXiv:hep-th/9607167 [hep-th]}
  \BibitemShut {NoStop}%
\bibitem [{\citenamefont {Bazhanov}\ \emph {et~al.}(1997)\citenamefont
  {Bazhanov}, \citenamefont {Lukyanov},\ and\ \citenamefont
  {Zamolodchikov}}]{Bazhanov:1996aq}%
  \BibitemOpen
  \bibfield  {author} {\bibinfo {author} {\bibfnamefont {V.~V.}\ \bibnamefont
  {Bazhanov}}, \bibinfo {author} {\bibfnamefont {S.~L.}\ \bibnamefont
  {Lukyanov}}, \ and\ \bibinfo {author} {\bibfnamefont {A.~B.}\ \bibnamefont
  {Zamolodchikov}},\ }\href {\doibase 10.1016/S0550-3213(97)00022-9} {\bibfield
   {journal} {\bibinfo  {journal} {Nucl. Phys.}\ }\textbf {\bibinfo {volume}
  {B489}},\ \bibinfo {pages} {487} (\bibinfo {year} {1997})},\ \Eprint
  {http://arxiv.org/abs/hep-th/9607099} {arXiv:hep-th/9607099 [hep-th]}
  \BibitemShut {NoStop}%
\bibitem [{\citenamefont {Caetano}()}]{Caetanounpublished}%
  \BibitemOpen
  \bibfield  {author} {\bibinfo {author} {\bibfnamefont {J.}~\bibnamefont
  {Caetano}},\ }\href@noop {} {\bibinfo  {journal} {unpublished}\ }\BibitemShut
  {NoStop}%
\bibitem [{\citenamefont {Grabner}\ \emph {et~al.}()\citenamefont {Grabner},
  \citenamefont {Gromov}, \citenamefont {Kazakov},\ and\ \citenamefont
  {Korchemsky}}]{GrabnerGromovKazakovKorchemsky}%
  \BibitemOpen
\bibfield  {journal} {  }\bibfield  {author} {\bibinfo {author} {\bibfnamefont
  {D.}~\bibnamefont {Grabner}}, \bibinfo {author} {\bibfnamefont
  {N.}~\bibnamefont {Gromov}}, \bibinfo {author} {\bibfnamefont
  {V.}~\bibnamefont {Kazakov}}, \ and\ \bibinfo {author} {\bibfnamefont
  {G.}~\bibnamefont {Korchemsky}},\ }\href@noop {} {\bibinfo  {journal} {to
  appear}\ }\BibitemShut {NoStop}%
\bibitem [{\citenamefont {Derkachov}\ \emph {et~al.}(2001)\citenamefont
  {Derkachov}, \citenamefont {Korchemsky},\ and\ \citenamefont
  {Manashov}}]{Derkachov:2001yn}%
  \BibitemOpen
\bibfield  {journal} {  }\bibfield  {author} {\bibinfo {author} {\bibfnamefont
  {S.~E.}\ \bibnamefont {Derkachov}}, \bibinfo {author} {\bibfnamefont {G.~P.}\
  \bibnamefont {Korchemsky}}, \ and\ \bibinfo {author} {\bibfnamefont {A.~N.}\
  \bibnamefont {Manashov}},\ }\href {\doibase 10.1016/S0550-3213(01)00457-6}
  {\bibfield  {journal} {\bibinfo  {journal} {Nucl. Phys.}\ }\textbf {\bibinfo
  {volume} {B617}},\ \bibinfo {pages} {375} (\bibinfo {year} {2001})},\ \Eprint
  {http://arxiv.org/abs/hep-th/0107193} {arXiv:hep-th/0107193 [hep-th]}
  \BibitemShut {NoStop}%
\bibitem [{\citenamefont {Gromov}\ \emph {et~al.}(2019)\citenamefont {Gromov},
  \citenamefont {Levkovich-Maslyuk}, \citenamefont {Ryan},\ and\ \citenamefont
  {Volin}}]{Gromov:2019wmz}%
  \BibitemOpen
  \bibfield  {author} {\bibinfo {author} {\bibfnamefont {N.}~\bibnamefont
  {Gromov}}, \bibinfo {author} {\bibfnamefont {F.}~\bibnamefont
  {Levkovich-Maslyuk}}, \bibinfo {author} {\bibfnamefont {P.}~\bibnamefont
  {Ryan}}, \ and\ \bibinfo {author} {\bibfnamefont {D.}~\bibnamefont {Volin}},\
  }\href@noop {} {\  (\bibinfo {year} {2019})},\ \Eprint
  {http://arxiv.org/abs/1910.13442} {arXiv:1910.13442 [hep-th]} \BibitemShut
  {NoStop}%
\bibitem [{\citenamefont {Basso}\ \emph {et~al.}(2018)\citenamefont {Basso},
  \citenamefont {Caetano},\ and\ \citenamefont {Fleury}}]{Basso:2018cvy}%
  \BibitemOpen
  \bibfield  {author} {\bibinfo {author} {\bibfnamefont {B.}~\bibnamefont
  {Basso}}, \bibinfo {author} {\bibfnamefont {J.}~\bibnamefont {Caetano}}, \
  and\ \bibinfo {author} {\bibfnamefont {T.}~\bibnamefont {Fleury}},\
  }\href@noop {} {\  (\bibinfo {year} {2018})},\ \Eprint
  {http://arxiv.org/abs/1812.09794} {arXiv:1812.09794 [hep-th]} \BibitemShut
  {NoStop}%
\bibitem [{\citenamefont {Ferrando}()}]{Ferrandoinprogress}%
  \BibitemOpen
  \bibfield  {author} {\bibinfo {author} {\bibfnamefont {G.}~\bibnamefont
  {Ferrando}},\ }\href@noop {} {\bibinfo  {journal} {in progress}\
  }\BibitemShut {NoStop}%
\end{thebibliography}%
\end{document}